\documentclass[%
 reprint,
 amsmath,amssymb,
 aps,
]{revtex4-2}

\usepackage{graphicx}
\usepackage{dcolumn}
\usepackage{bm}

\usepackage{color}     
\usepackage{amsmath}
\usepackage{multirow}  

\begin{document}

\preprint{APS/123-QED}

\title{Stellar weak rates of the $rp$-process waiting points: Effects of strong magnetic fields}

\author{Qi-Ye Hu}%
\affiliation{School of Physical Science and Technology, Southwest University, Chongqing 400715, China}%
\author{Long-Jun Wang}
\email{longjun@swu.edu.cn}
\affiliation{School of Physical Science and Technology, Southwest University, Chongqing 400715, China} 
\author{Yang Sun}
\affiliation{School of Physics and Astronomy, Shanghai Jiao Tong University, Shanghai 200240, China}%

\date{\today}

\begin{abstract}
  Incorporating  microscopic nuclear-structure information into the discussion of bulk properties of astronomical objects such as neutron stars has always been a challenging issue in interdisciplinary nuclear astrophysics. Using the $rp$-process nucleosynthesis as an example, we studied the effective stellar $\beta^+$ and electron capture (EC) rates of eight waiting-point (WP) nuclei with realistic stellar conditions and presence of strong magnetic fields. The relevant nuclear transition strengths are provided by the projected shell model. We have found that, on average, due to the magnetic field effect, the $\beta^+$ and EC rates can increase by more than an order of magnitude for all combinations of density and temperature, as well as in each of WPs studied. We relate the onset field strength, at which the weak rates begin to increase, to nuclear structure quantities, $Q$ value or electronic chemical potential $\mu_e$. The enhanced weak rates may change considerably the lifetime of WPs, thereby modifying the current understanding of the $rp$-process.
   
\end{abstract}

\maketitle



Intense magnetic fields on surfaces of neutron stars can significantly alter the physical properties of bulk matter through modifying the characteristics of the atomic nuclei exposed to the fields. Studying the effects of such magnetic fields is necessary, for example, to explain  detailed spectrum of neutrino emission during the cooling process of neutron stars \cite{Famiano_2022_ApJ}, which has to do with finding an explanation for the cooling of magnetars, a peculiar type of slow-rotating neutron star with a super-strong magnetic field on its surface. 

On the other hand, neutron stars provide possible sites for stellar nucleosynthesis of heavy elements through rapid neutron-capture ($r$-) process and rapid proton-capture ($rp$-) process \cite{r_process_RMP_2021, rp_process_Schatz_1998}. Study of effects of magnetic fields on stellar nuclear weak-interaction processes is crucial for understanding the neutron-star structure and stellar nucleosynthesis \cite{Lai_ApJ_1991, Lai_RMP_2001, Zhang_Jie_2006_CPC}. Recently, Refs. \cite{Famiano_2020_ApJ, Kajino_2014_ApJ, Famiano_PRD_2020} intensively discussed the effects of strong magnetic fields on the $\beta^-$-decay rates for $r$-process neuron-rich nuclei. 

The $rp$-process takes place in accreting neutron stars, by thermally unstable nuclear burning in low mass X-ray binaries (LMXBs), and/or by thermally stable burning in high mass X-ray binaries (HMXBs) \cite{rp_process_Schatz_1998, Schatz_1999_ApJ, Schatz_2001_PRL_end_of_rp}. The former includes type I X-ray bursts and rarely pulsation with weak magnetic fields $B \approx 10^{8-10}$ G, while the latter is usually referred to as X-ray pulsars  with very strong magnetic fields $B \approx 10^{12-13}$ G \cite{Coburn_2002_ApJ, Revnivtsev_2015, Cheatham_thesis_2017} or even higher ($\approx 10^{14}$ G) \cite{Science_2017_NGC5907}, depending on modeling.

\begin{figure}
\begin{center}
  \includegraphics[width=0.49\textwidth]{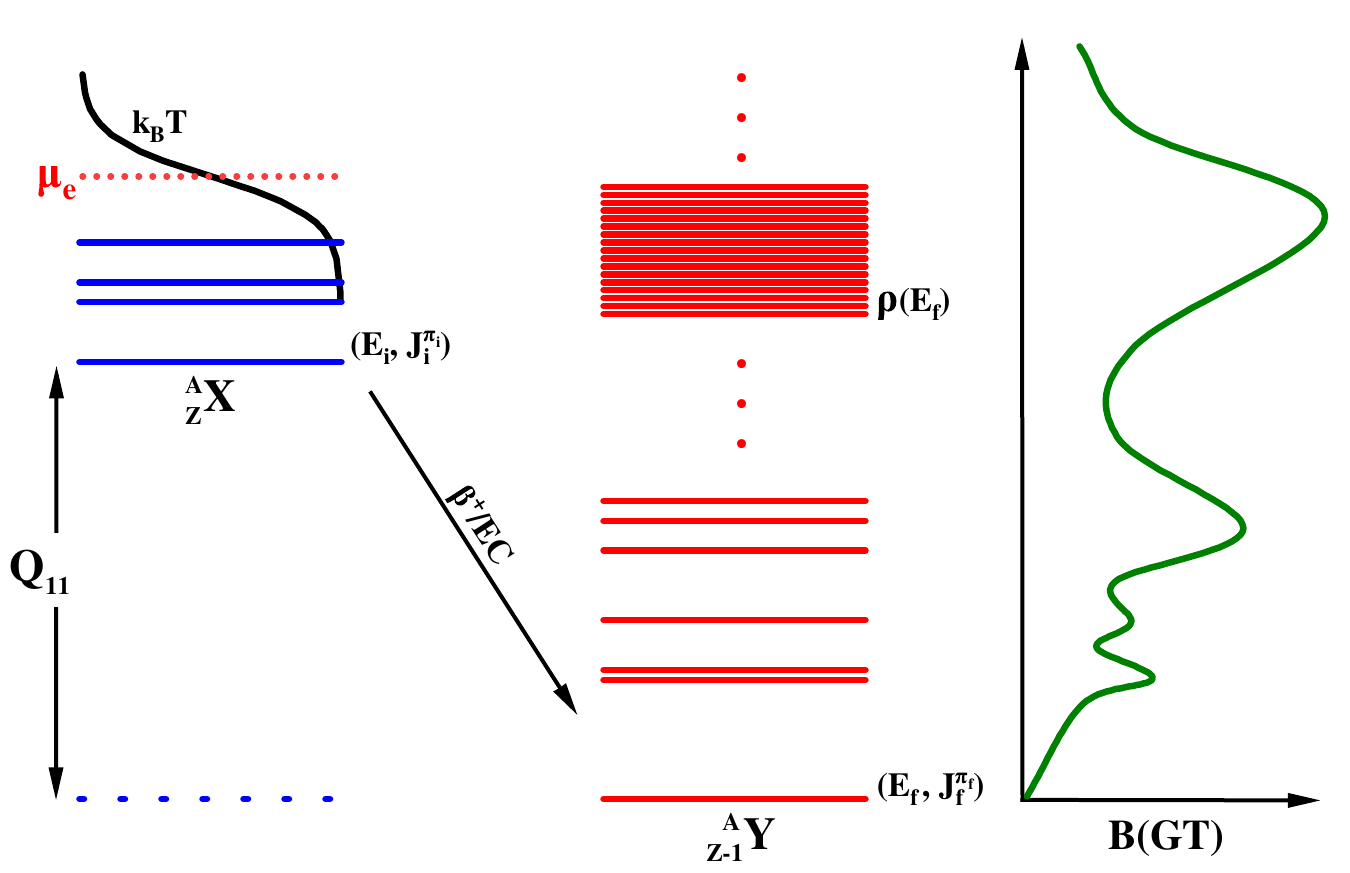}
  \caption{\label{fig:schematic} A schematic diagram for weak decay scheme with a proton-rich nucleus $^A_Z$X at stellar conditions. Both $\beta^+$ decay and EC that connect the daughter nucleus $_{Z-1}^{\ \ A}$Y are considered. The rapid increase of nuclear level density $\rho(E_x)$ with excitation energy $E_x$ is illustrated. } 
\end{center}
\end{figure}

In the present Letter, we discuss the influence of stellar weak-interaction rates for nuclei near the $N=Z$ line with presence of strong magnetic fields, which is relevant for the $rp$-process. Our study distinguishes it from previous ones in two important aspects. Firstly, we treat both $\beta^+$ decay and electron capture (EC) channels on an equal footing, in which the latter, notably, has positive $Q$ values for proton-rich nuclei (see Fig. \ref{fig:schematic} for schematic illustration), in contrast to negative ones for $r$-process neuron-rich nuclei, and therefore, their roles can be significantly different as the environment changes with presence of strong magnetic fields. Secondly, unlike the prevalent one-transition assumption (see, for example, Ref. \cite{schatz2014nature}), we consider realistic weak-transitions from the ground state and thermally-populated low-lying states of parent nuclei ($^A_Z$X in Fig. \ref{fig:schematic}) to the ground state and all excited states up to 10 MeV of excitation in daughter nuclei ($_{Z-1}^{\ \ A}$Y in Fig. \ref{fig:schematic}). Detailed nuclear properties such as the $Q$ values and shape factors for the study are properly treated by the state-of-the-art shell model.

The $rp$-process waiting points (WPs) refer to those $N=Z$ nuclei on the $rp$-path, where further proton capture is suspended and the process is paused to wait for slow $\beta^+$ decays. The effective half-life of the WP nuclei is crucial for determining the path of the $rp$-process and the final element abundances in the corresponding stellar environments \cite{Schatz_2001_PRL_end_of_rp}. In this Letter we consider eight most important WP nuclei, $^{64}$Ge, $^{68}$Se, $^{72}$Kr, $^{76}$Sr, $^{80}$Zr, $^{84}$Mo, $^{88}$Ru and $^{92}$Pd, to study effects of strong magnetic fields on the stellar $\beta^+$ and EC rates. 


The weak-interaction rates in stellar environments with strong magnetic fields are expressed as
\begin{eqnarray} \label{eq.total_lambda}
  \lambda^\alpha = \sum_{if} \frac{(2J_{i}+1)e^{-E_{i}/(k_{B}T)}}{G(Z,A,T)} \lambda^\alpha_{if},
\end{eqnarray}
where $\alpha$ labels the $\beta^+$ or EC channel, and $k_{B}$ is the Boltzmann constant and $T$ the environment temperature. The summation in (\ref{eq.total_lambda}) runs over the initial ($i$) and final ($f$) states of the parent and daughter nuclei, respectively, with each having a definite spin-parity $J_i^{\pi_i}$ or $J_f^{\pi_f}$ and an excitation energy $E_i$ or $ E_f$. $G(Z, A, T) = \sum_i (2J_i + 1)\text{exp}(-E_{i}/(k_{B}T))$ is the partition function. 

With presence of strong magnetic fields, the electron distribution is anisotropic in the momentum space: electron motion in the direction perpendicular to magnetic fields is quantized into Landau levels. Individual rates $\lambda^\alpha_{if}$ read as \cite{Lai_ApJ_1991, Xiao_Wang_PRC_2024}
\begin{subequations} \label{eq.lambda_if}
\begin{align} 
  \lambda^{\beta^+}_{if(B)}  =& \frac{\ln 2}{K} \frac{B^\ast}{2} \sum_{n=0}^{N_{\text{max}}} (2-\delta_{n0}) \int_{0}^{ p_{znu} } C(W_n) (Q_{if}-W_n)^2 \nonumber \\
                              & \qquad \times F_0(-Z+1, W_n) (1-S_p(W_n)) dp_z,  \\
  \lambda^{\text{EC}}_{if(B)} =& \frac{\ln 2}{K} \frac{B^\ast}{2} \sum_{n=0}^{N_{\text{max}}} (2-\delta_{n0}) \int_{ p_{znl} }^{ \infty } C(W_n) (Q_{if}+W_n)^2 \nonumber \\
                              & \qquad \times F_0(Z, W_n) S_e (W_n) dp_z, 
\end{align}
\end{subequations}
where $K=6144 \pm 2$ s is the constant adopted from \cite{Hardy_2009_PRC}, and $B^\ast \equiv B/B_c$ the dimensionless magnetic field  strength in units of the critical magnetic field $B_c = m_e^2 c^3 / e \hbar \approx 4.414 \times 10^{13}$ G \cite{Lai_ApJ_1991}. The summations in (\ref{eq.lambda_if}) run over the Landau levels labeled by $n$ with spin degeneracy $(2-\delta_{n0})$ \cite{Lai_ApJ_1991, Lai_RMP_2001} up to the maximum number $N_{\text{max}} = (Q_{if}^2 - 1) / 2B^\ast$, where $Q_{if} = (M_p - M_d + E_i -E_f ) / m_e c^2 $ is (dimensionless) available total energy for leptons in individual transitions, and $M_p (M_d)$ the nuclear mass of parent (daughter) nucleus. The integral variable $p_z$ is the momentum component in the direction of the magnetic field and the integral limits are $p_{znu} = \sqrt{ Q^2_{if} - 1 - 2nB^\ast }$, and
\begin{eqnarray}
  p_{znl} = 
  \left\{ \begin{array}{cl} 
      \sqrt{ Q^2_{if} - 1 - 2nB^\ast }  & \text{for } Q_{if} < -\sqrt{1+2nB^\ast}, \\ 
                                 0      & \text{for } Q_{if} \geqslant -\sqrt{1+2nB^\ast}, 
  \end{array} \right. 
\end{eqnarray}
where $p_{znu}$ ($p_{znl}$), the integral upper (lower) limit of $p_z$, corresponds to the maximum (minimum) energy of electrons allowed by the phase space in $\beta^+$ (EC). $W_n = \sqrt{p^2_z + 1 + 2nB^\ast}$ is the electron energy. $F_0$ is the Fermi function that accounts for the Coulomb distortion of the electron wave function near the nucleus \cite{Fuller1980, Fermi_func_1983}. 

In Eq. (\ref{eq.lambda_if}), the electron and positron distribution functions follow the Fermi-Dirac distribution,
\begin{eqnarray} \label{eq.Se}
  S_{e/p} (W_n) = \frac{1}{\text{exp}[(W_n \mp \mu_e) / k_B T] + 1} ,
\end{eqnarray}
where the electron chemical potential $\mu_{e}$, with  presence of magnetic fields, is determined by
\begin{eqnarray} \label{eq.mue}
  \rho Y_e = \frac{B^\ast}{2\pi^2 N_A \lambdabar_e^3} \sum_{n=0}^{N_{ \text{lim} }} (2-\delta_{n0}) \int_{0}^{\infty} (S_e - S_p) dp_z ,
\end{eqnarray}
where $\rho Y_e$ labels the electron density, $N_A$ is the Avogadro's number, and $\lambdabar_e$ the  reduced Compton wavelength of electron. $N_{\text{lim}} \approx \mu_e^2 / 2B^\ast$ is the effective limit of the number of Laudau levels. 

$C(W_n)$ in Eq. (\ref{eq.lambda_if}), the shape factor \cite{Zhi_FF_PRC_2013, Xiao_Wang_PRC_2024} for individual nuclear transitions, contains details in nuclear structure, and therefore, plays a decisive role in the discussion. As it is very difficult to measure shape factors experimentally, especially for exotic nuclei, their values must rely on calculations. In the present work, we adopt the projected shell model (PSM) \cite{LJWang_2014_PRC_Rapid, LJWang_2016_PRC, LJWang_2018_PRC_GT, LJWang_PLB_2020_ec, LJWang_2021_PRL, LJWang_2021_PRC_93Nb, BLWang_1stF_2024} that contains a sufficiently large configuration space, which is essential for obtaining stellar weak rates for highly-excited nuclear states \cite{LJWang_2021_PRL}. As previous studies of the $rp$-process \cite{P.sarriguren2001NPA, P.sarriguren2005EPJA, Sarriguren2009PLB, sarriguren2012JP, nabi2012AASS, nabi2016beta, Nabi2017AASS, A.petrovici2011PPNP, A.petrovici2015EPJA, Petrovici_2019_PRC, R.Lau2018NPA, R.Lau2020MN}, we consider only allowed transitions (with $\pi_i \pi_f = +1$). The reduced Gamow-Teller transition strengths, $B(\text{GT}^+)$, are taken from the previous PSM calculation \cite{ZRChen_PLB2024}. 

\begin{figure}
\begin{center}
  \includegraphics[width=0.49\textwidth]{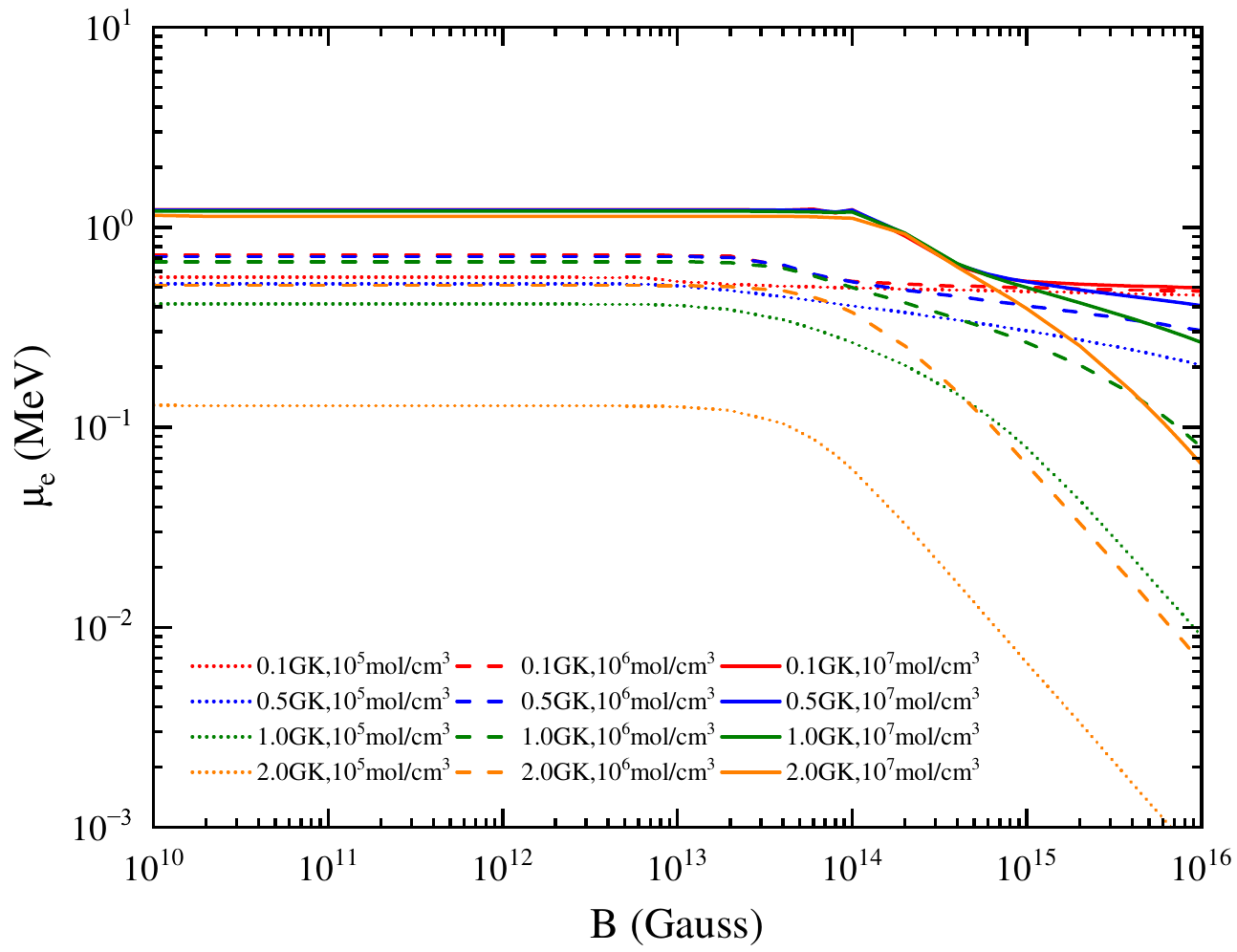}
  \caption{\label{fig:mue} Calculated chemical potential of electrons, $\mu_e$, for different parameter combinations of density (in mol/cm$^3$) and temperature (in GK), as a function of magnetic field strength. }
\end{center}
\end{figure}

In Fig. \ref{fig:mue}, we first show the electron chemical potential, $\mu_e$, as a function of magnetic field strength $B$, for different stellar densities and temperatures. As can be seen, $\mu_e$ is sensitive to electron density: For weak $B$'s, the higher the density, the larger the $\mu_e$. It depends less on temperature. With increasing $B$, $\mu_e$ remains constant for low $B$'s. It starts to decrease when $B$ reaches a certain strong value and decreases more rapidly with higher $T$. 

\begin{figure*}
\begin{center}
  \includegraphics[width=0.98\textwidth]{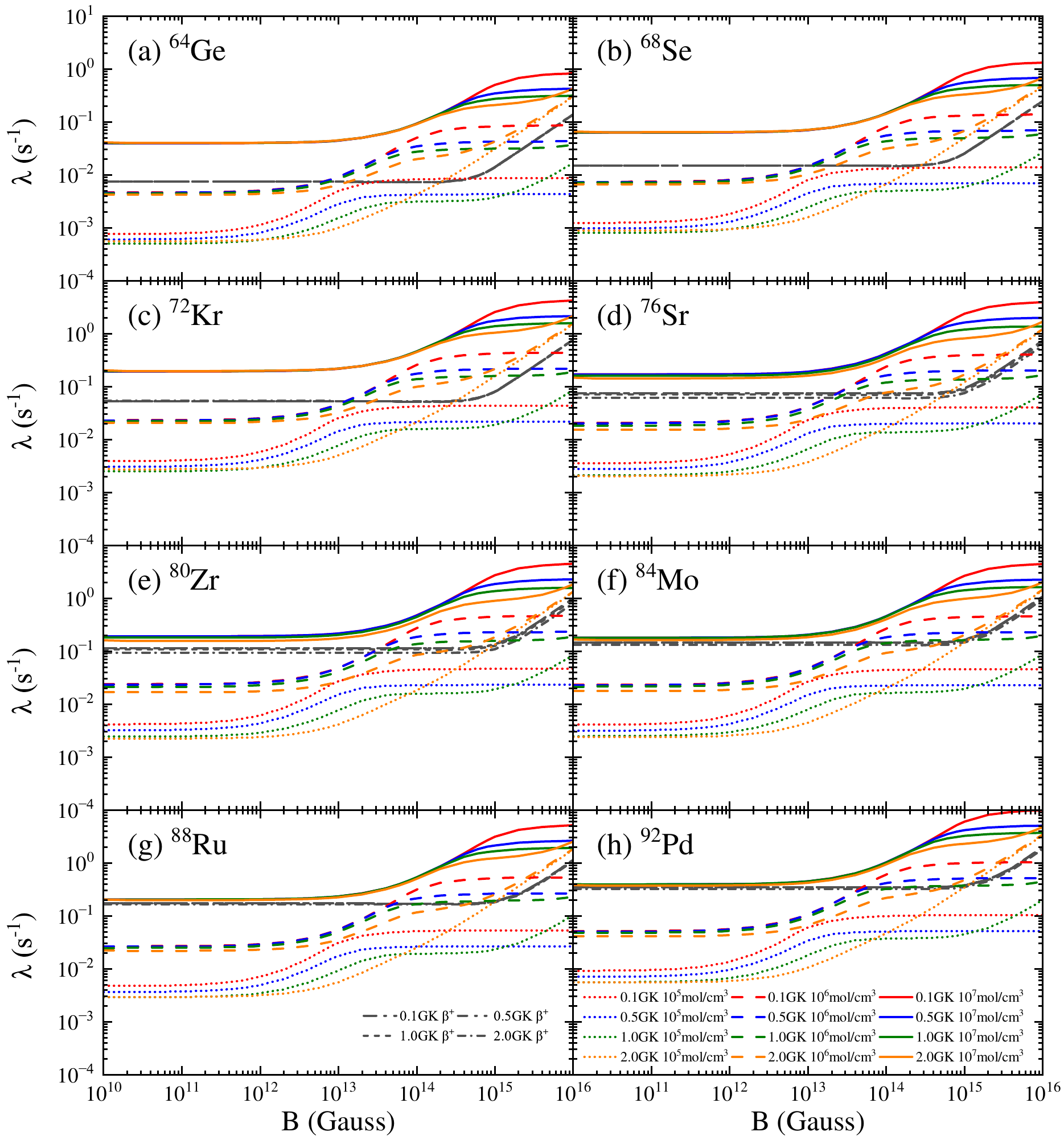}
  \caption{\label{fig:lambda} Calculated $\beta^+$-decay rates (not colored) and EC rates (colored) of the eight $rp$-process waiting point nuclei for different parameter combinations of density (in mol/cm$^3$) and temperature (in GK), as a function of the magnetic field. Curve parameters for $\beta^+$-decay and EC rates are explained in Panel (g) and (h), respectively. Note that the $\beta^+$-decay rates are not dependent of density while the EC rates depend sensitively on both density and temperature. }
\end{center}
\end{figure*}





These results can be understood as follows. For given density $\rho Y_e$ and temperature $T$, although the degeneracy of state density for each Landau level or cylinder (which is proportional to $B^\ast$) increases rapidly with increasing $B$, more and more high-lying Landau levels are repelled out of the Fermi sphere due to the increasing energy separation between them, so that the effective number of Landau levels within the Fermi sphere (i.e. $N_{\text{lim}}+1$), decreases rapidly for low $B$ case. The balance of these two aspects keeps the Fermi sphere, $\mu_e$, unchanged. This corresponds to the region of constant $\mu_e$ in Fig. \ref{fig:mue}. When $B$ increases to beyond a critical value, further reduction in the effective number of Landau levels within the Fermi sphere is no longer possible to balance the rapid increase in the degeneracy of state density, the Fermi sphere, thus $\mu_e$, has to shrink to keep the density $\rho Y_e$ on the left-hand side of Eq. (\ref{eq.mue}) invariable. The decrease in $\mu_e$ is illustrated in Fig. \ref{fig:mue} for the strong-$B$ region. The dependence of $\mu_e$ on $B$ for different $T$ is determined by the $T$ dependent on $S_{e/p}$ in Eq. (\ref{eq.Se}). 

Using $C(W_n)$ and $\mu_e$, $\beta^+$ and EC rates of the eight $rp$-process WP nuclei are evaluated for different stellar densities, temperatures $T$ and magnetic field $B$ as shown in Fig. \ref{fig:lambda}. Noticeably, the response of the two rates to the environment is very different. The stellar $\beta^+$ rates are not sensitive to the stellar density and $T$. In contrast, the EC rates vary sensitively with the environment variables. 

With increasing $B$, $\beta^+$ rates remain unchanged for a wide range of field strength, but rise rapidly for $B \gtrsim 10^{15}$ G. The field strength at the onset of the sudden change in $\lambda^{\beta^+}$, $B_\text{onset}$,  which is about $4 \times 10^{14}$ G ($10^{15}$ G) for $^{64}$Ge ($^{92}$Pd) (see Fig. \ref{fig:lambda} (a) and (h)), is found to correlate with the $Q$-value (such as $Q_{11}$ between the ground states shown in Fig. \ref{fig:schematic}). By analyzing the integral of $\lambda^{\beta^+}_{if(B)}$ in Eq. (\ref{eq.lambda_if}a), it is found that the total $\lambda^{\beta^+}$ in Eq. (\ref{eq.total_lambda}) is often dominated by a few $\lambda^{\beta^+}_{if(B)}$ with very large (or the largest) $Q_{if}$ value(s). To simplify the discussion, we suppose that $\lambda^{\beta^+}$ is dominated by one single rate $\lambda^{\beta^+}_{11(B)}$ (the transition between ground states with $Q_{if} = Q_{11}$). When the field is weak, the decrease in the number of Landau levels (i.e. $N_{\text{max}} + 1$) offsets the increasing degeneracy of state density due to increasing $B$, keeping $\lambda^{\beta^+}$ a  constant. When the field is sufficiently strong so that the number of Landau levels is reduced to a few (for example, $N_{\text{max}} \approx 5$), the balance cannot be kept further, and $\lambda^{\beta^+}$ begins to increase linearly with $B$, as seen in Fig. \ref{fig:lambda}. The onset field strength, $B_\text{onset}$, can be estimated by $N_{\text{max}} \approx Q_{11}^2 / 2B^\ast \approx 5$, so that $B_\text{onset} \approx Q_{11}^2 \times 4.4 \times 10^{12}$ G. For $^{64}$Ge and $^{92}$Pd with $Q_{11} \approx 8$ and 16 \cite{ZRChen_PLB2024}, one has $B_\text{onset} \approx 3 \times 10^{14}$ G and $10^{15}$ G, respectively, as seen in Fig. \ref{fig:lambda} (a) and (h). 

The above estimate for $B_\text{onset}$ may well be applied also to stellar $\beta^-$-decay rates, $\lambda^{\beta^-}$, for neutron-rich nuclei in the $r$ process. There, $\lambda^{\beta^-}$ is determined by $Q_{if} - \mu_e$, where $Q_{if}$ is the possible maximum $Q$-value for transitions with large $B(\text{GT}^-)$ and $\mu_e$ depends on stellar density. For cases of $Q_{11} \gg \mu_e$, the influence on $\mu_e$ with increasing $B$ is negligible, and $\lambda^{\beta^-}$ should arise linearly with $B$ beyond $B_\text{onset} \approx Q_{11}^2 \times 4.4 \times 10^{12}$ G. However, for $Q_{if} \gtrsim \mu_e$, a case-by-case analysis may be indispensable when studying the effect of magnetic fields on $\lambda^{\beta^-}$.

The stellar EC rate, $\lambda^{\text{EC}}$, behaves very differently compared with $\beta$ decay rates for increasing magnetic fields. By comparing the integrands and integral limits in Eqs. (\ref{eq.lambda_if}b) and (\ref{eq.lambda_if}a), one can see that not only the transitions to the ground state and very low-lying states with large $Q_{if}$, but also the transitions with many smaller $Q_{if}$ to {\it highly-excited states} may contribute effectively to the total $\lambda^{\text{EC}}$. Detailed contribution from different transitions depends on the distribution of the shape factor $C(W_n)$ (i.e. on the $B(\text{GT})$ distribution if only allowed transitions are considered), as seen from Eq. (\ref{eq.lambda_if}b) and Fig. \ref{fig:schematic}.

Furthermore, because individual $\lambda^{\text{EC}}_{if(B)}$ is sensitive to  $\mu_e$ through  $S_e(W_n)$ and its phase space is proportional to $Q_{if} + \mu_e + k_B T$ (see Fig. \ref{fig:schematic}), the total $\lambda^{\text{EC}}$ shows remarkable sensitivity to stellar density and temperature. More interestingly, for specific $B$ and $\mu_e$, $S_e(W_n)$ decreases with increasing $n$, which suggests that the effective number of Landau levels, $N_{\text{max}}+1$, that contribute to $\lambda^{\text{EC}}_{if(B)}$, should be determined by $S_e(W_n)$ with $2 N_{\text{max}} B^\ast \approx \mu_e^2$ for EC, not simply by a single $Q_{11}$ as in $\beta$-decay cases. 

When the field is weak, the decrease of $N_{\text{max}}$ can offset the effect of increasing $B$, as discussed above, and $\lambda^{\text{EC}}$ remains unchanged. When the field becomes sufficiently strong so that $N_{\text{max}}$ is reduced to a small number, $\lambda^{\text{EC}}$ begins to increase. The onset field strength for the increase of $\lambda^{\text{EC}}$ can be estimated by $B_{\text{onset}} \approx \mu_e^2 \times 4.4 \times 10^{12}$ G. Using the constant $\mu_e$ values at the weak magnetic fields in Fig. \ref{fig:mue}, one can estimate that $B_{\text{onset}} \approx 10^{12}$ G, $5 \times 10^{12}$ G, and $ 2\times 10^{13}$ G for $\rho Y_e = 10^5$, $10^6$ and $10^7$ mol/cm$^3$, respectively, consistent with Fig. \ref{fig:lambda}. When $B \gtrsim 10^{14}$ G, $\mu_e$ begins to decrease rapidly (see Fig. \ref{fig:mue}), so that the decreasing $S_e(W_n)$ may balance the effect of increasing $B$, leading to plateau structures of $\lambda^{\text{EC}}$ in many cases, as seen from Fig. \ref{fig:lambda}. 

Based on the above discussion, we can draw a general conclusion for the results of $\lambda^{\beta^+}$ and $\lambda^{\text{EC}}$. The presence of strong magnetic fields leads to significant enhancements in the weak rates, with details depending on environment parameters and field strength. On average, an increase in the rates by more than an order of magnitude is evident for all WP nuclei on the $rp$-process path. The global trend with increasing $B$ seems similar in the eight panels of Fig. \ref{fig:lambda}. However, the competition between $\lambda^{\text{EC}}$ and $\lambda^{\beta^+}$ with increasing fields is determined by structural quantities such as $B(\text{GT})$ distributions and $Q$ values. For spherical or near-spherical nuclei (such as $^{92}$Pd) with large $Q$ values and much concentrated $B(\text{GT})$ distributions \cite{ZRChen_PLB2024, Lv_Cui_Juan_2022_PRC}, $\lambda^{\beta^+}$ dominates the total rates until a very strong field ($\approx 10^{14}$ G for $^{92}$Pd) comes into play. For deformed nuclei (such as $^{64}$Ge and $^{68}$Se) with smaller $Q$ values and much fragmented $B(\text{GT})$ distributions \cite{ZRChen_PLB2024, Lv_Cui_Juan_2022_PRC}, the role of $\lambda^{\text{EC}}$ may become comparable with, or even overtakes $\lambda^{\beta^+}$ for weak fields (for $^{64}$Ge with $\rho Y_e = 10^6$ mol/cm$^3$, $\lambda^{\text{EC}}$ becomes dominant once $B \gtrsim 10^{13}$ G). Such enhanced rates can in general reduce lifetimes of the WPs significantly.





To summarize, we studied the effect of strong magnetic fields on the stellar weak interaction rates of the $rp$-process waiting point nuclei. As  distinctive features of our method, we (1) considered both $\beta^+$ decay and electron capture channels, and (2) considered state-to-state transitions between many excited levels of the parent and daughter nuclei. Specifically, through a detailed analysis of the eight waiting points in the $rp$-process, we found that because of the magnetic-field effect, $\lambda^{\beta^+}$ and $\lambda^{\text{EC}}$ are significantly enhanced. Perhaps it is never noticed before, but due to modifications in the electron distribution, the impact is much more pronounced on electron capture, either through  rearrangement of electron Landau levels or change in the electron phase space. Depending on actual astrophysical environments, enhanced weak rates may impact the $rp$-process by modifying the lifetime of the waiting-point nuclei, or conversely, the possible $rp$-process sites and stellar conditions may be better constrained with the help of our works. The present method can be extended to general studies of $r$- and $rp$-process problems with magnetic fields, which are in progress.    


\begin{acknowledgments}
  We thank J. M. Dong for motivating us to study the effects of magnetic field on stellar nuclear weak decays, and H. Schatz for valuable communications. This work is supported by the National Natural Science Foundation of China (Grant Nos. 12275225, 12235003). 
\end{acknowledgments}






\begin{thebibliography}{42}%
\makeatletter
\providecommand \@ifxundefined [1]{%
 \@ifx{#1\undefined}
}%
\providecommand \@ifnum [1]{%
 \ifnum #1\expandafter \@firstoftwo
 \else \expandafter \@secondoftwo
 \fi
}%
\providecommand \@ifx [1]{%
 \ifx #1\expandafter \@firstoftwo
 \else \expandafter \@secondoftwo
 \fi
}%
\providecommand \natexlab [1]{#1}%
\providecommand \enquote  [1]{``#1''}%
\providecommand \bibnamefont  [1]{#1}%
\providecommand \bibfnamefont [1]{#1}%
\providecommand \citenamefont [1]{#1}%
\providecommand \href@noop [0]{\@secondoftwo}%
\providecommand \href [0]{\begingroup \@sanitize@url \@href}%
\providecommand \@href[1]{\@@startlink{#1}\@@href}%
\providecommand \@@href[1]{\endgroup#1\@@endlink}%
\providecommand \@sanitize@url [0]{\catcode `\\12\catcode `\$12\catcode
  `\&12\catcode `\#12\catcode `\^12\catcode `\_12\catcode `\%12\relax}%
\providecommand \@@startlink[1]{}%
\providecommand \@@endlink[0]{}%
\providecommand \url  [0]{\begingroup\@sanitize@url \@url }%
\providecommand \@url [1]{\endgroup\@href {#1}{\urlprefix }}%
\providecommand \urlprefix  [0]{URL }%
\providecommand \Eprint [0]{\href }%
\providecommand \doibase [0]{https://doi.org/}%
\providecommand \selectlanguage [0]{\@gobble}%
\providecommand \bibinfo  [0]{\@secondoftwo}%
\providecommand \bibfield  [0]{\@secondoftwo}%
\providecommand \translation [1]{[#1]}%
\providecommand \BibitemOpen [0]{}%
\providecommand \bibitemStop [0]{}%
\providecommand \bibitemNoStop [0]{.\EOS\space}%
\providecommand \EOS [0]{\spacefactor3000\relax}%
\providecommand \BibitemShut  [1]{\csname bibitem#1\endcsname}%
\let\auto@bib@innerbib\@empty
\bibitem [{\citenamefont {Famiano}\ \emph {et~al.}(2022)\citenamefont
  {Famiano}, \citenamefont {Mathews}, \citenamefont {Balantekin}, \citenamefont
  {Kajino}, \citenamefont {Kusakabe},\ and\ \citenamefont
  {Mori}}]{Famiano_2022_ApJ}%
  \BibitemOpen
  \bibfield  {author} {\bibinfo {author} {\bibfnamefont {M.~A.}\ \bibnamefont
  {Famiano}}, \bibinfo {author} {\bibfnamefont {G.}~\bibnamefont {Mathews}},
  \bibinfo {author} {\bibfnamefont {A.~B.}\ \bibnamefont {Balantekin}},
  \bibinfo {author} {\bibfnamefont {T.}~\bibnamefont {Kajino}}, \bibinfo
  {author} {\bibfnamefont {M.}~\bibnamefont {Kusakabe}},\ and\ \bibinfo
  {author} {\bibfnamefont {K.}~\bibnamefont {Mori}},\ }\bibfield  {title}
  {\bibinfo {title} {Evolution of urca pairs in the crusts of highly magnetized
  neutron stars},\ }\href {https://doi.org/10.3847/1538-4357/ac9bf3} {\bibfield
   {journal} {\bibinfo  {journal} {The Astrophysical Journal}\ }\textbf
  {\bibinfo {volume} {940}},\ \bibinfo {pages} {108} (\bibinfo {year}
  {2022})}\BibitemShut {NoStop}%
\bibitem [{\citenamefont {Cowan}\ \emph {et~al.}(2021)\citenamefont {Cowan},
  \citenamefont {Sneden}, \citenamefont {Lawler}, \citenamefont {Aprahamian},
  \citenamefont {Wiescher}, \citenamefont {Langanke}, \citenamefont
  {Mart\'{\i}nez-Pinedo},\ and\ \citenamefont
  {Thielemann}}]{r_process_RMP_2021}%
  \BibitemOpen
  \bibfield  {author} {\bibinfo {author} {\bibfnamefont {J.~J.}\ \bibnamefont
  {Cowan}}, \bibinfo {author} {\bibfnamefont {C.}~\bibnamefont {Sneden}},
  \bibinfo {author} {\bibfnamefont {J.~E.}\ \bibnamefont {Lawler}}, \bibinfo
  {author} {\bibfnamefont {A.}~\bibnamefont {Aprahamian}}, \bibinfo {author}
  {\bibfnamefont {M.}~\bibnamefont {Wiescher}}, \bibinfo {author}
  {\bibfnamefont {K.}~\bibnamefont {Langanke}}, \bibinfo {author}
  {\bibfnamefont {G.}~\bibnamefont {Mart\'{\i}nez-Pinedo}},\ and\ \bibinfo
  {author} {\bibfnamefont {F.-K.}\ \bibnamefont {Thielemann}},\ }\bibfield
  {title} {\bibinfo {title} {Origin of the heaviest elements: The rapid
  neutron-capture process},\ }\href
  {https://doi.org/10.1103/RevModPhys.93.015002} {\bibfield  {journal}
  {\bibinfo  {journal} {Rev. Mod. Phys.}\ }\textbf {\bibinfo {volume} {93}},\
  \bibinfo {pages} {015002} (\bibinfo {year} {2021})}\BibitemShut {NoStop}%
\bibitem [{\citenamefont {Schatz}\ \emph {et~al.}(1998)\citenamefont {Schatz},
  \citenamefont {Aprahamian}, \citenamefont {Görres}, \citenamefont
  {Wiescher}, \citenamefont {Rauscher}, \citenamefont {Rembges}, \citenamefont
  {Thielemann}, \citenamefont {Pfeiffer}, \citenamefont {Möller},
  \citenamefont {Kratz}, \citenamefont {Herndl}, \citenamefont {Brown},\ and\
  \citenamefont {Rebel}}]{rp_process_Schatz_1998}%
  \BibitemOpen
  \bibfield  {author} {\bibinfo {author} {\bibfnamefont {H.}~\bibnamefont
  {Schatz}}, \bibinfo {author} {\bibfnamefont {A.}~\bibnamefont {Aprahamian}},
  \bibinfo {author} {\bibfnamefont {J.}~\bibnamefont {Görres}}, \bibinfo
  {author} {\bibfnamefont {M.}~\bibnamefont {Wiescher}}, \bibinfo {author}
  {\bibfnamefont {T.}~\bibnamefont {Rauscher}}, \bibinfo {author}
  {\bibfnamefont {J.}~\bibnamefont {Rembges}}, \bibinfo {author} {\bibfnamefont
  {F.-K.}\ \bibnamefont {Thielemann}}, \bibinfo {author} {\bibfnamefont
  {B.}~\bibnamefont {Pfeiffer}}, \bibinfo {author} {\bibfnamefont
  {P.}~\bibnamefont {Möller}}, \bibinfo {author} {\bibfnamefont {K.-L.}\
  \bibnamefont {Kratz}}, \bibinfo {author} {\bibfnamefont {H.}~\bibnamefont
  {Herndl}}, \bibinfo {author} {\bibfnamefont {B.}~\bibnamefont {Brown}},\ and\
  \bibinfo {author} {\bibfnamefont {H.}~\bibnamefont {Rebel}},\ }\bibfield
  {title} {\bibinfo {title} {rp-process nucleosynthesis at extreme temperature
  and density conditions},\ }\href
  {https://doi.org/https://doi.org/10.1016/S0370-1573(97)00048-3} {\bibfield
  {journal} {\bibinfo  {journal} {Physics Reports}\ }\textbf {\bibinfo {volume}
  {294}},\ \bibinfo {pages} {167} (\bibinfo {year} {1998})}\BibitemShut
  {NoStop}%
\bibitem [{\citenamefont {Lai}\ and\ \citenamefont
  {Shapiro}(1991)}]{Lai_ApJ_1991}%
  \BibitemOpen
  \bibfield  {author} {\bibinfo {author} {\bibfnamefont {D.}~\bibnamefont
  {Lai}}\ and\ \bibinfo {author} {\bibfnamefont {S.~L.}\ \bibnamefont
  {Shapiro}},\ }\bibfield  {title} {\bibinfo {title} {Cold equation of state in
  a strong magnetic field: Effects of inverse beta-decay},\ }\href
  {http://dx.doi.org/10.1086/170831} {\bibfield  {journal} {\bibinfo  {journal}
  {Astrophysical Journal}\ }\textbf {\bibinfo {volume} {383}},\ \bibinfo
  {pages} {745} (\bibinfo {year} {1991})}\BibitemShut {NoStop}%
\bibitem [{\citenamefont {Lai}(2001)}]{Lai_RMP_2001}%
  \BibitemOpen
  \bibfield  {author} {\bibinfo {author} {\bibfnamefont {D.}~\bibnamefont
  {Lai}},\ }\bibfield  {title} {\bibinfo {title} {Matter in strong magnetic
  fields},\ }\href {https://doi.org/10.1103/RevModPhys.73.629} {\bibfield
  {journal} {\bibinfo  {journal} {Rev. Mod. Phys.}\ }\textbf {\bibinfo {volume}
  {73}},\ \bibinfo {pages} {629} (\bibinfo {year} {2001})}\BibitemShut
  {NoStop}%
\bibitem [{\citenamefont {Zhang}\ \emph {et~al.}(2006)\citenamefont {Zhang},
  \citenamefont {Liu},\ and\ \citenamefont {Luo}}]{Zhang_Jie_2006_CPC}%
  \BibitemOpen
  \bibfield  {author} {\bibinfo {author} {\bibfnamefont {J.}~\bibnamefont
  {Zhang}}, \bibinfo {author} {\bibfnamefont {M.-Q.}\ \bibnamefont {Liu}},\
  and\ \bibinfo {author} {\bibfnamefont {Z.-Q.}\ \bibnamefont {Luo}},\
  }\bibfield  {title} {\bibinfo {title} {Influence of strong magnetic field on
  $\beta$ decay in the crusts of neutron stars},\ }\href
  {https://doi.org/10.1088/1009-1963/15/7/016} {\bibfield  {journal} {\bibinfo
  {journal} {Chinese Physics}\ }\textbf {\bibinfo {volume} {15}},\ \bibinfo
  {pages} {1477} (\bibinfo {year} {2006})}\BibitemShut {NoStop}%
\bibitem [{\citenamefont {Famiano}\ \emph {et~al.}(2020)\citenamefont
  {Famiano}, \citenamefont {Balantekin}, \citenamefont {Kajino}, \citenamefont
  {Kusakabe}, \citenamefont {Mori},\ and\ \citenamefont
  {Luo}}]{Famiano_2020_ApJ}%
  \BibitemOpen
  \bibfield  {author} {\bibinfo {author} {\bibfnamefont {M.}~\bibnamefont
  {Famiano}}, \bibinfo {author} {\bibfnamefont {A.~B.}\ \bibnamefont
  {Balantekin}}, \bibinfo {author} {\bibfnamefont {T.}~\bibnamefont {Kajino}},
  \bibinfo {author} {\bibfnamefont {M.}~\bibnamefont {Kusakabe}}, \bibinfo
  {author} {\bibfnamefont {K.}~\bibnamefont {Mori}},\ and\ \bibinfo {author}
  {\bibfnamefont {Y.}~\bibnamefont {Luo}},\ }\bibfield  {title} {\bibinfo
  {title} {Nuclear reaction screening, weak interactions, and r-process
  nucleosynthesis in high magnetic fields},\ }\href
  {https://doi.org/10.3847/1538-4357/aba04d} {\bibfield  {journal} {\bibinfo
  {journal} {The Astrophysical Journal}\ }\textbf {\bibinfo {volume} {898}},\
  \bibinfo {pages} {163} (\bibinfo {year} {2020})}\BibitemShut {NoStop}%
\bibitem [{\citenamefont {Kajino}\ \emph {et~al.}(2014)\citenamefont {Kajino},
  \citenamefont {Tokuhisa}, \citenamefont {Mathews}, \citenamefont {Yoshida},\
  and\ \citenamefont {Famiano}}]{Kajino_2014_ApJ}%
  \BibitemOpen
  \bibfield  {author} {\bibinfo {author} {\bibfnamefont {T.}~\bibnamefont
  {Kajino}}, \bibinfo {author} {\bibfnamefont {A.}~\bibnamefont {Tokuhisa}},
  \bibinfo {author} {\bibfnamefont {G.~J.}\ \bibnamefont {Mathews}}, \bibinfo
  {author} {\bibfnamefont {T.}~\bibnamefont {Yoshida}},\ and\ \bibinfo {author}
  {\bibfnamefont {M.~A.}\ \bibnamefont {Famiano}},\ }\bibfield  {title}
  {\bibinfo {title} {Ultra high-energy neutrinos via heavy-meson synchrotron
  emission in strong magnetic fields},\ }\href
  {https://doi.org/10.1088/0004-637X/782/2/70} {\bibfield  {journal} {\bibinfo
  {journal} {The Astrophysical Journal}\ }\textbf {\bibinfo {volume} {782}},\
  \bibinfo {pages} {70} (\bibinfo {year} {2014})}\BibitemShut {NoStop}%
\bibitem [{\citenamefont {Luo}\ \emph {et~al.}(2020)\citenamefont {Luo},
  \citenamefont {Famiano}, \citenamefont {Kajino}, \citenamefont {Kusakabe},\
  and\ \citenamefont {Balantekin}}]{Famiano_PRD_2020}%
  \BibitemOpen
  \bibfield  {author} {\bibinfo {author} {\bibfnamefont {Y.}~\bibnamefont
  {Luo}}, \bibinfo {author} {\bibfnamefont {M.~A.}\ \bibnamefont {Famiano}},
  \bibinfo {author} {\bibfnamefont {T.}~\bibnamefont {Kajino}}, \bibinfo
  {author} {\bibfnamefont {M.}~\bibnamefont {Kusakabe}},\ and\ \bibinfo
  {author} {\bibfnamefont {A.~B.}\ \bibnamefont {Balantekin}},\ }\bibfield
  {title} {\bibinfo {title} {Screening corrections to electron capture rates
  and resulting constraints on primordial magnetic fields},\ }\href
  {https://doi.org/10.1103/PhysRevD.101.083010} {\bibfield  {journal} {\bibinfo
   {journal} {Phys. Rev. D}\ }\textbf {\bibinfo {volume} {101}},\ \bibinfo
  {pages} {083010} (\bibinfo {year} {2020})}\BibitemShut {NoStop}%
\bibitem [{\citenamefont {Schatz}\ \emph {et~al.}(1999)\citenamefont {Schatz},
  \citenamefont {Bildsten}, \citenamefont {Cumming},\ and\ \citenamefont
  {Wiescher}}]{Schatz_1999_ApJ}%
  \BibitemOpen
  \bibfield  {author} {\bibinfo {author} {\bibfnamefont {H.}~\bibnamefont
  {Schatz}}, \bibinfo {author} {\bibfnamefont {L.}~\bibnamefont {Bildsten}},
  \bibinfo {author} {\bibfnamefont {A.}~\bibnamefont {Cumming}},\ and\ \bibinfo
  {author} {\bibfnamefont {M.}~\bibnamefont {Wiescher}},\ }\bibfield  {title}
  {\bibinfo {title} {The rapid proton process ashes from stable nuclear burning
  on an accreting neutron star},\ }\href {https://doi.org/10.1086/307837}
  {\bibfield  {journal} {\bibinfo  {journal} {The Astrophysical Journal}\
  }\textbf {\bibinfo {volume} {524}},\ \bibinfo {pages} {1014} (\bibinfo {year}
  {1999})}\BibitemShut {NoStop}%
\bibitem [{\citenamefont {Schatz}\ \emph {et~al.}(2001)\citenamefont {Schatz},
  \citenamefont {Aprahamian}, \citenamefont {Barnard}, \citenamefont
  {Bildsten}, \citenamefont {Cumming}, \citenamefont {Ouellette}, \citenamefont
  {Rauscher}, \citenamefont {Thielemann},\ and\ \citenamefont
  {Wiescher}}]{Schatz_2001_PRL_end_of_rp}%
  \BibitemOpen
  \bibfield  {author} {\bibinfo {author} {\bibfnamefont {H.}~\bibnamefont
  {Schatz}}, \bibinfo {author} {\bibfnamefont {A.}~\bibnamefont {Aprahamian}},
  \bibinfo {author} {\bibfnamefont {V.}~\bibnamefont {Barnard}}, \bibinfo
  {author} {\bibfnamefont {L.}~\bibnamefont {Bildsten}}, \bibinfo {author}
  {\bibfnamefont {A.}~\bibnamefont {Cumming}}, \bibinfo {author} {\bibfnamefont
  {M.}~\bibnamefont {Ouellette}}, \bibinfo {author} {\bibfnamefont
  {T.}~\bibnamefont {Rauscher}}, \bibinfo {author} {\bibfnamefont {F.-K.}\
  \bibnamefont {Thielemann}},\ and\ \bibinfo {author} {\bibfnamefont
  {M.}~\bibnamefont {Wiescher}},\ }\bibfield  {title} {\bibinfo {title} {End
  point of the $\mathit{rp}$ process on accreting neutron stars},\ }\href
  {https://doi.org/10.1103/PhysRevLett.86.3471} {\bibfield  {journal} {\bibinfo
   {journal} {Phys. Rev. Lett.}\ }\textbf {\bibinfo {volume} {86}},\ \bibinfo
  {pages} {3471} (\bibinfo {year} {2001})}\BibitemShut {NoStop}%
\bibitem [{\citenamefont {Coburn}\ \emph {et~al.}(2002)\citenamefont {Coburn},
  \citenamefont {Heindl}, \citenamefont {Rothschild}, \citenamefont {Gruber},
  \citenamefont {Kreykenbohm}, \citenamefont {Wilms}, \citenamefont
  {Kretschmar},\ and\ \citenamefont {Staubert}}]{Coburn_2002_ApJ}%
  \BibitemOpen
  \bibfield  {author} {\bibinfo {author} {\bibfnamefont {W.}~\bibnamefont
  {Coburn}}, \bibinfo {author} {\bibfnamefont {W.~A.}\ \bibnamefont {Heindl}},
  \bibinfo {author} {\bibfnamefont {R.~E.}\ \bibnamefont {Rothschild}},
  \bibinfo {author} {\bibfnamefont {D.~E.}\ \bibnamefont {Gruber}}, \bibinfo
  {author} {\bibfnamefont {I.}~\bibnamefont {Kreykenbohm}}, \bibinfo {author}
  {\bibfnamefont {J.}~\bibnamefont {Wilms}}, \bibinfo {author} {\bibfnamefont
  {P.}~\bibnamefont {Kretschmar}},\ and\ \bibinfo {author} {\bibfnamefont
  {R.}~\bibnamefont {Staubert}},\ }\bibfield  {title} {\bibinfo {title}
  {Magnetic fields of accreting x-ray pulsars with the rossi x-ray timing
  explorer},\ }\href {https://doi.org/10.1086/343033} {\bibfield  {journal}
  {\bibinfo  {journal} {The Astrophysical Journal}\ }\textbf {\bibinfo {volume}
  {580}},\ \bibinfo {pages} {394} (\bibinfo {year} {2002})}\BibitemShut
  {NoStop}%
\bibitem [{\citenamefont {Revnivtsev}\ and\ \citenamefont
  {Mereghetti}(2015)}]{Revnivtsev_2015}%
  \BibitemOpen
  \bibfield  {author} {\bibinfo {author} {\bibfnamefont {M.}~\bibnamefont
  {Revnivtsev}}\ and\ \bibinfo {author} {\bibfnamefont {S.}~\bibnamefont
  {Mereghetti}},\ }\bibfield  {title} {\bibinfo {title} {Magnetic fields of
  neutron stars in x-ray binaries},\ }\href
  {https://doi.org/110.1007/s11214-014-0123-x} {\bibfield  {journal} {\bibinfo
  {journal} {Space Science Reviews}\ }\textbf {\bibinfo {volume} {191}},\
  \bibinfo {pages} {293} (\bibinfo {year} {2015})}\BibitemShut {NoStop}%
\bibitem [{\citenamefont {Cheatham}(2017)}]{Cheatham_thesis_2017}%
  \BibitemOpen
  \bibfield  {author} {\bibinfo {author} {\bibfnamefont {D.~M.}\ \bibnamefont
  {Cheatham}},\ }\emph {\bibinfo {title} {Shedding new light on magnetic
  accretion: A comprehensive study of the X-ray emission in accreting
  pulsars}},\ \href@noop {} {\bibinfo {type} {{Ph.D.} thesis}},\ \bibinfo
  {school} {University of Maryland} (\bibinfo {year} {2017})\BibitemShut
  {NoStop}%
\bibitem [{\citenamefont {Israel}\ \emph {et~al.}(2017)\citenamefont {Israel},
  \citenamefont {Belfiore}, \citenamefont {Stella}, \citenamefont {Esposito},
  \citenamefont {Casella}, \citenamefont {Luca}, \citenamefont {Marelli},
  \citenamefont {Papitto}, \citenamefont {Perri}, \citenamefont {Puccetti},
  \citenamefont {Castillo}, \citenamefont {Salvetti}, \citenamefont {Tiengo},
  \citenamefont {Zampieri}, \citenamefont {D’Agostino}, \citenamefont
  {Greiner}, \citenamefont {Haberl}, \citenamefont {Novara}, \citenamefont
  {Salvaterra}, \citenamefont {Turolla}, \citenamefont {Watson}, \citenamefont
  {Wilms},\ and\ \citenamefont {Wolter}}]{Science_2017_NGC5907}%
  \BibitemOpen
  \bibfield  {author} {\bibinfo {author} {\bibfnamefont {G.~L.}\ \bibnamefont
  {Israel}}, \bibinfo {author} {\bibfnamefont {A.}~\bibnamefont {Belfiore}},
  \bibinfo {author} {\bibfnamefont {L.}~\bibnamefont {Stella}}, \bibinfo
  {author} {\bibfnamefont {P.}~\bibnamefont {Esposito}}, \bibinfo {author}
  {\bibfnamefont {P.}~\bibnamefont {Casella}}, \bibinfo {author} {\bibfnamefont
  {A.~D.}\ \bibnamefont {Luca}}, \bibinfo {author} {\bibfnamefont
  {M.}~\bibnamefont {Marelli}}, \bibinfo {author} {\bibfnamefont
  {A.}~\bibnamefont {Papitto}}, \bibinfo {author} {\bibfnamefont
  {M.}~\bibnamefont {Perri}}, \bibinfo {author} {\bibfnamefont
  {S.}~\bibnamefont {Puccetti}}, \bibinfo {author} {\bibfnamefont {G.~A.~R.}\
  \bibnamefont {Castillo}}, \bibinfo {author} {\bibfnamefont {D.}~\bibnamefont
  {Salvetti}}, \bibinfo {author} {\bibfnamefont {A.}~\bibnamefont {Tiengo}},
  \bibinfo {author} {\bibfnamefont {L.}~\bibnamefont {Zampieri}}, \bibinfo
  {author} {\bibfnamefont {D.}~\bibnamefont {D’Agostino}}, \bibinfo {author}
  {\bibfnamefont {J.}~\bibnamefont {Greiner}}, \bibinfo {author} {\bibfnamefont
  {F.}~\bibnamefont {Haberl}}, \bibinfo {author} {\bibfnamefont
  {G.}~\bibnamefont {Novara}}, \bibinfo {author} {\bibfnamefont
  {R.}~\bibnamefont {Salvaterra}}, \bibinfo {author} {\bibfnamefont
  {R.}~\bibnamefont {Turolla}}, \bibinfo {author} {\bibfnamefont
  {M.}~\bibnamefont {Watson}}, \bibinfo {author} {\bibfnamefont
  {J.}~\bibnamefont {Wilms}},\ and\ \bibinfo {author} {\bibfnamefont
  {A.}~\bibnamefont {Wolter}},\ }\bibfield  {title} {\bibinfo {title} {An
  accreting pulsar with extreme properties drives an ultraluminous x-ray source
  in ngc 5907},\ }\href {https://doi.org/10.1126/science.aai8635} {\bibfield
  {journal} {\bibinfo  {journal} {Science}\ }\textbf {\bibinfo {volume}
  {355}},\ \bibinfo {pages} {817} (\bibinfo {year} {2017})},\ \Eprint
  {https://arxiv.org/abs/https://www.science.org/doi/pdf/10.1126/science.aai8635}
  {https://www.science.org/doi/pdf/10.1126/science.aai8635} \BibitemShut
  {NoStop}%
\bibitem [{\citenamefont {Schatz}\ \emph {et~al.}(2014)\citenamefont {Schatz},
  \citenamefont {Gupta}, \citenamefont {M\"oller}, \citenamefont {Beard},
  \citenamefont {Brown}, \citenamefont {Deibel}, \citenamefont {Gasques},
  \citenamefont {Hix}, \citenamefont {Keek}, \citenamefont {Lau}, \citenamefont
  {Steiner},\ and\ \citenamefont {Wiescher}}]{schatz2014nature}%
  \BibitemOpen
  \bibfield  {author} {\bibinfo {author} {\bibfnamefont {H.}~\bibnamefont
  {Schatz}}, \bibinfo {author} {\bibfnamefont {S.}~\bibnamefont {Gupta}},
  \bibinfo {author} {\bibfnamefont {P.}~\bibnamefont {M\"oller}}, \bibinfo
  {author} {\bibfnamefont {M.}~\bibnamefont {Beard}}, \bibinfo {author}
  {\bibfnamefont {E.~F.}\ \bibnamefont {Brown}}, \bibinfo {author}
  {\bibfnamefont {A.~T.}\ \bibnamefont {Deibel}}, \bibinfo {author}
  {\bibfnamefont {L.~R.}\ \bibnamefont {Gasques}}, \bibinfo {author}
  {\bibfnamefont {W.~R.}\ \bibnamefont {Hix}}, \bibinfo {author} {\bibfnamefont
  {L.}~\bibnamefont {Keek}}, \bibinfo {author} {\bibfnamefont {R.}~\bibnamefont
  {Lau}}, \bibinfo {author} {\bibfnamefont {A.~W.}\ \bibnamefont {Steiner}},\
  and\ \bibinfo {author} {\bibfnamefont {M.}~\bibnamefont {Wiescher}},\
  }\bibfield  {title} {\bibinfo {title} {Strong neutrino cooling by cycles of
  electron capture and $\beta^-$ decay in neutron star crusts},\ }\href
  {https://doi.org/10.1038/nature12757} {\bibfield  {journal} {\bibinfo
  {journal} {Nature}\ }\textbf {\bibinfo {volume} {505}},\ \bibinfo {pages}
  {62} (\bibinfo {year} {2014})}\BibitemShut {NoStop}%
\bibitem [{\citenamefont {Xiao}\ \emph {et~al.}(2024)\citenamefont {Xiao},
  \citenamefont {Wang},\ and\ \citenamefont {Wang}}]{Xiao_Wang_PRC_2024}%
  \BibitemOpen
  \bibfield  {author} {\bibinfo {author} {\bibfnamefont {Y.}~\bibnamefont
  {Xiao}}, \bibinfo {author} {\bibfnamefont {B.-L.}\ \bibnamefont {Wang}},\
  and\ \bibinfo {author} {\bibfnamefont {L.-J.}\ \bibnamefont {Wang}},\
  }\bibfield  {title} {\bibinfo {title} {Stellar $\ensuremath{\beta}$-decay
  rate of the $s$-process branching-point $^{204}\mathrm{Tl}$: Forbidden
  transitions},\ }\href {https://doi.org/10.1103/PhysRevC.110.015806}
  {\bibfield  {journal} {\bibinfo  {journal} {Phys. Rev. C}\ }\textbf {\bibinfo
  {volume} {110}},\ \bibinfo {pages} {015806} (\bibinfo {year}
  {2024})}\BibitemShut {NoStop}%
\bibitem [{\citenamefont {Hardy}\ and\ \citenamefont
  {Towner}(2009)}]{Hardy_2009_PRC}%
  \BibitemOpen
  \bibfield  {author} {\bibinfo {author} {\bibfnamefont {J.~C.}\ \bibnamefont
  {Hardy}}\ and\ \bibinfo {author} {\bibfnamefont {I.~S.}\ \bibnamefont
  {Towner}},\ }\bibfield  {title} {\bibinfo {title} {Superallowed
  ${0}^{+}\ensuremath{\rightarrow}{0}^{+}$ nuclear $\ensuremath{\beta}$ decays:
  A new survey with precision tests of the conserved vector current hypothesis
  and the standard model},\ }\href {https://doi.org/10.1103/PhysRevC.79.055502}
  {\bibfield  {journal} {\bibinfo  {journal} {Phys. Rev. C}\ }\textbf {\bibinfo
  {volume} {79}},\ \bibinfo {pages} {055502} (\bibinfo {year}
  {2009})}\BibitemShut {NoStop}%
\bibitem [{\citenamefont {Fuller}\ \emph {et~al.}(1980)\citenamefont {Fuller},
  \citenamefont {Fowler},\ and\ \citenamefont {Newman}}]{Fuller1980}%
  \BibitemOpen
  \bibfield  {author} {\bibinfo {author} {\bibfnamefont {G.~M.}\ \bibnamefont
  {Fuller}}, \bibinfo {author} {\bibfnamefont {W.~A.}\ \bibnamefont {Fowler}},\
  and\ \bibinfo {author} {\bibfnamefont {M.~J.}\ \bibnamefont {Newman}},\
  }\bibfield  {title} {\bibinfo {title} {Stellar weak-interaction rates for
  sd-shell nuclei. i. nuclear matrix element systematics with application to
  $^{26}$\text{Al} and selected nuclei of imprtance to the supernova problem},\
  }\href {https://doi.org/Doi 10.1086/190657} {\bibfield  {journal} {\bibinfo
  {journal} {Astrophys. J. (Suppl.)}\ }\textbf {\bibinfo {volume} {42}},\
  \bibinfo {pages} {447} (\bibinfo {year} {1980})}\BibitemShut {NoStop}%
\bibitem [{\citenamefont {Schenter}\ and\ \citenamefont
  {Vogel}(1983)}]{Fermi_func_1983}%
  \BibitemOpen
  \bibfield  {author} {\bibinfo {author} {\bibfnamefont {G.~K.}\ \bibnamefont
  {Schenter}}\ and\ \bibinfo {author} {\bibfnamefont {P.}~\bibnamefont
  {Vogel}},\ }\bibfield  {title} {\bibinfo {title} {A simple approximation of
  the fermi function in nuclear beta decay},\ }\href
  {https://doi.org/10.13182/NSE83-A17574} {\bibfield  {journal} {\bibinfo
  {journal} {Nuclear Science and Engineering}\ }\textbf {\bibinfo {volume}
  {83}},\ \bibinfo {pages} {393} (\bibinfo {year} {1983})},\ \Eprint
  {https://arxiv.org/abs/https://doi.org/10.13182/NSE83-A17574}
  {https://doi.org/10.13182/NSE83-A17574} \BibitemShut {NoStop}%
\bibitem [{\citenamefont {Zhi}\ \emph {et~al.}(2013)\citenamefont {Zhi},
  \citenamefont {Caurier}, \citenamefont {Cuenca-Garc\'{\i}a}, \citenamefont
  {Langanke}, \citenamefont {Mart\'{\i}nez-Pinedo},\ and\ \citenamefont
  {Sieja}}]{Zhi_FF_PRC_2013}%
  \BibitemOpen
  \bibfield  {author} {\bibinfo {author} {\bibfnamefont {Q.}~\bibnamefont
  {Zhi}}, \bibinfo {author} {\bibfnamefont {E.}~\bibnamefont {Caurier}},
  \bibinfo {author} {\bibfnamefont {J.~J.}\ \bibnamefont {Cuenca-Garc\'{\i}a}},
  \bibinfo {author} {\bibfnamefont {K.}~\bibnamefont {Langanke}}, \bibinfo
  {author} {\bibfnamefont {G.}~\bibnamefont {Mart\'{\i}nez-Pinedo}},\ and\
  \bibinfo {author} {\bibfnamefont {K.}~\bibnamefont {Sieja}},\ }\bibfield
  {title} {\bibinfo {title} {Shell-model half-lives including first-forbidden
  contributions for $r$-process waiting-point nuclei},\ }\href
  {https://doi.org/10.1103/PhysRevC.87.025803} {\bibfield  {journal} {\bibinfo
  {journal} {Phys. Rev. C}\ }\textbf {\bibinfo {volume} {87}},\ \bibinfo
  {pages} {025803} (\bibinfo {year} {2013})}\BibitemShut {NoStop}%
\bibitem [{\citenamefont {Wang}\ \emph {et~al.}(2014)\citenamefont {Wang},
  \citenamefont {Chen}, \citenamefont {Mizusaki}, \citenamefont {Oi},\ and\
  \citenamefont {Sun}}]{LJWang_2014_PRC_Rapid}%
  \BibitemOpen
  \bibfield  {author} {\bibinfo {author} {\bibfnamefont {L.-J.}\ \bibnamefont
  {Wang}}, \bibinfo {author} {\bibfnamefont {F.-Q.}\ \bibnamefont {Chen}},
  \bibinfo {author} {\bibfnamefont {T.}~\bibnamefont {Mizusaki}}, \bibinfo
  {author} {\bibfnamefont {M.}~\bibnamefont {Oi}},\ and\ \bibinfo {author}
  {\bibfnamefont {Y.}~\bibnamefont {Sun}},\ }\bibfield  {title} {\bibinfo
  {title} {Toward extremes of angular momentum: Application of the pfaffian
  algorithm in realistic calculations},\ }\href
  {https://doi.org/10.1103/PhysRevC.90.011303} {\bibfield  {journal} {\bibinfo
  {journal} {Phys. Rev. C}\ }\textbf {\bibinfo {volume} {90}},\ \bibinfo
  {pages} {011303(R)} (\bibinfo {year} {2014})}\BibitemShut {NoStop}%
\bibitem [{\citenamefont {Wang}\ \emph {et~al.}(2016)\citenamefont {Wang},
  \citenamefont {Sun}, \citenamefont {Mizusaki}, \citenamefont {Oi},\ and\
  \citenamefont {Ghorui}}]{LJWang_2016_PRC}%
  \BibitemOpen
  \bibfield  {author} {\bibinfo {author} {\bibfnamefont {L.-J.}\ \bibnamefont
  {Wang}}, \bibinfo {author} {\bibfnamefont {Y.}~\bibnamefont {Sun}}, \bibinfo
  {author} {\bibfnamefont {T.}~\bibnamefont {Mizusaki}}, \bibinfo {author}
  {\bibfnamefont {M.}~\bibnamefont {Oi}},\ and\ \bibinfo {author}
  {\bibfnamefont {S.~K.}\ \bibnamefont {Ghorui}},\ }\bibfield  {title}
  {\bibinfo {title} {Reduction of collectivity at very high spins in
  $^{134}\mathrm{Nd}$: Expanding the projected-shell-model basis up to
  10-quasiparticle states},\ }\href
  {https://doi.org/10.1103/PhysRevC.93.034322} {\bibfield  {journal} {\bibinfo
  {journal} {Phys. Rev. C}\ }\textbf {\bibinfo {volume} {93}},\ \bibinfo
  {pages} {034322} (\bibinfo {year} {2016})}\BibitemShut {NoStop}%
\bibitem [{\citenamefont {Wang}\ \emph {et~al.}(2018)\citenamefont {Wang},
  \citenamefont {Sun},\ and\ \citenamefont {Ghorui}}]{LJWang_2018_PRC_GT}%
  \BibitemOpen
  \bibfield  {author} {\bibinfo {author} {\bibfnamefont {L.-J.}\ \bibnamefont
  {Wang}}, \bibinfo {author} {\bibfnamefont {Y.}~\bibnamefont {Sun}},\ and\
  \bibinfo {author} {\bibfnamefont {S.~K.}\ \bibnamefont {Ghorui}},\ }\bibfield
   {title} {\bibinfo {title} {Shell-model method for gamow-teller transitions
  in heavy deformed odd-mass nuclei},\ }\href
  {https://doi.org/10.1103/PhysRevC.97.044302} {\bibfield  {journal} {\bibinfo
  {journal} {Phys. Rev. C}\ }\textbf {\bibinfo {volume} {97}},\ \bibinfo
  {pages} {044302} (\bibinfo {year} {2018})}\BibitemShut {NoStop}%
\bibitem [{\citenamefont {Tan}\ \emph {et~al.}(2020)\citenamefont {Tan},
  \citenamefont {Liu}, \citenamefont {Wang}, \citenamefont {Li},\ and\
  \citenamefont {Sun}}]{LJWang_PLB_2020_ec}%
  \BibitemOpen
  \bibfield  {author} {\bibinfo {author} {\bibfnamefont {L.}~\bibnamefont
  {Tan}}, \bibinfo {author} {\bibfnamefont {Y.-X.}\ \bibnamefont {Liu}},
  \bibinfo {author} {\bibfnamefont {L.-J.}\ \bibnamefont {Wang}}, \bibinfo
  {author} {\bibfnamefont {Z.}~\bibnamefont {Li}},\ and\ \bibinfo {author}
  {\bibfnamefont {Y.}~\bibnamefont {Sun}},\ }\bibfield  {title} {\bibinfo
  {title} {A novel method for stellar electron-capture rates of excited nuclear
  states},\ }\href
  {https://doi.org/https://doi.org/10.1016/j.physletb.2020.135432} {\bibfield
  {journal} {\bibinfo  {journal} {Phys. Lett. B}\ }\textbf {\bibinfo {volume}
  {805}},\ \bibinfo {pages} {135432} (\bibinfo {year} {2020})}\BibitemShut
  {NoStop}%
\bibitem [{\citenamefont {Wang}\ \emph
  {et~al.}(2021{\natexlab{a}})\citenamefont {Wang}, \citenamefont {Tan},
  \citenamefont {Li}, \citenamefont {Misch},\ and\ \citenamefont
  {Sun}}]{LJWang_2021_PRL}%
  \BibitemOpen
  \bibfield  {author} {\bibinfo {author} {\bibfnamefont {L.-J.}\ \bibnamefont
  {Wang}}, \bibinfo {author} {\bibfnamefont {L.}~\bibnamefont {Tan}}, \bibinfo
  {author} {\bibfnamefont {Z.}~\bibnamefont {Li}}, \bibinfo {author}
  {\bibfnamefont {G.~W.}\ \bibnamefont {Misch}},\ and\ \bibinfo {author}
  {\bibfnamefont {Y.}~\bibnamefont {Sun}},\ }\bibfield  {title} {\bibinfo
  {title} {Urca cooling in neutron star crusts and oceans: Effects of nuclear
  excitations},\ }\href {https://doi.org/10.1103/PhysRevLett.127.172702}
  {\bibfield  {journal} {\bibinfo  {journal} {Phys. Rev. Lett.}\ }\textbf
  {\bibinfo {volume} {127}},\ \bibinfo {pages} {172702} (\bibinfo {year}
  {2021}{\natexlab{a}})}\BibitemShut {NoStop}%
\bibitem [{\citenamefont {Wang}\ \emph
  {et~al.}(2021{\natexlab{b}})\citenamefont {Wang}, \citenamefont {Tan},
  \citenamefont {Li}, \citenamefont {Gao},\ and\ \citenamefont
  {Sun}}]{LJWang_2021_PRC_93Nb}%
  \BibitemOpen
  \bibfield  {author} {\bibinfo {author} {\bibfnamefont {L.-J.}\ \bibnamefont
  {Wang}}, \bibinfo {author} {\bibfnamefont {L.}~\bibnamefont {Tan}}, \bibinfo
  {author} {\bibfnamefont {Z.}~\bibnamefont {Li}}, \bibinfo {author}
  {\bibfnamefont {B.}~\bibnamefont {Gao}},\ and\ \bibinfo {author}
  {\bibfnamefont {Y.}~\bibnamefont {Sun}},\ }\bibfield  {title} {\bibinfo
  {title} {Description of $^{93}\mathrm{Nb}$ stellar electron-capture rates by
  the projected shell model},\ }\href
  {https://doi.org/10.1103/PhysRevC.104.064323} {\bibfield  {journal} {\bibinfo
   {journal} {Phys. Rev. C}\ }\textbf {\bibinfo {volume} {104}},\ \bibinfo
  {pages} {064323} (\bibinfo {year} {2021}{\natexlab{b}})}\BibitemShut
  {NoStop}%
\bibitem [{\citenamefont {Wang}\ and\ \citenamefont
  {Wang}(2024)}]{BLWang_1stF_2024}%
  \BibitemOpen
  \bibfield  {author} {\bibinfo {author} {\bibfnamefont {B.-L.}\ \bibnamefont
  {Wang}}\ and\ \bibinfo {author} {\bibfnamefont {L.-J.}\ \bibnamefont
  {Wang}},\ }\bibfield  {title} {\bibinfo {title} {First-forbidden transition
  of nuclear $\beta$ decay by projected shell model},\ }\href
  {https://doi.org/https://doi.org/10.1016/j.physletb.2024.138515} {\bibfield
  {journal} {\bibinfo  {journal} {Phys. Lett. B}\ }\textbf {\bibinfo {volume}
  {850}},\ \bibinfo {pages} {138515} (\bibinfo {year} {2024})}\BibitemShut
  {NoStop}%
\bibitem [{\citenamefont {Sarriguren}\ \emph {et~al.}(2001)\citenamefont
  {Sarriguren}, \citenamefont {{Moya de Guerra}},\ and\ \citenamefont
  {Escuderos}}]{P.sarriguren2001NPA}%
  \BibitemOpen
  \bibfield  {author} {\bibinfo {author} {\bibfnamefont {P.}~\bibnamefont
  {Sarriguren}}, \bibinfo {author} {\bibfnamefont {E.}~\bibnamefont {{Moya de
  Guerra}}},\ and\ \bibinfo {author} {\bibfnamefont {A.}~\bibnamefont
  {Escuderos}},\ }\bibfield  {title} {\bibinfo {title} {Spin–isospin
  excitations and $\beta+$/ec half-lives of medium-mass deformed nuclei},\
  }\href {https://doi.org/https://doi.org/10.1016/S0375-9474(01)00565-6}
  {\bibfield  {journal} {\bibinfo  {journal} {Nucl. Phys. A}\ }\textbf
  {\bibinfo {volume} {691}},\ \bibinfo {pages} {631} (\bibinfo {year}
  {2001})}\BibitemShut {NoStop}%
\bibitem [{\citenamefont {Sarriguren}\ \emph {et~al.}(2005)\citenamefont
  {Sarriguren}, \citenamefont {Alvarez-Rodr{\i}guez},\ and\ \citenamefont
  {Moya~de Guerra}}]{P.sarriguren2005EPJA}%
  \BibitemOpen
  \bibfield  {author} {\bibinfo {author} {\bibfnamefont {P.}~\bibnamefont
  {Sarriguren}}, \bibinfo {author} {\bibfnamefont {R.}~\bibnamefont
  {Alvarez-Rodr{\i}guez}},\ and\ \bibinfo {author} {\bibfnamefont
  {E.}~\bibnamefont {Moya~de Guerra}},\ }\bibfield  {title} {\bibinfo {title}
  {Half-lives of rp-process waiting point nuclei},\ }\href@noop {} {\bibfield
  {journal} {\bibinfo  {journal} {Eur. Phys. J. A}\ }\textbf {\bibinfo {volume}
  {24}},\ \bibinfo {pages} {193} (\bibinfo {year} {2005})}\BibitemShut
  {NoStop}%
\bibitem [{\citenamefont {Sarriguren}(2009)}]{Sarriguren2009PLB}%
  \BibitemOpen
  \bibfield  {author} {\bibinfo {author} {\bibfnamefont {P.}~\bibnamefont
  {Sarriguren}},\ }\bibfield  {title} {\bibinfo {title} {Weak interaction rates
  for kr and sr waiting-point nuclei under rp-process conditions},\ }\href
  {https://doi.org/https://doi.org/10.1016/j.physletb.2009.09.046} {\bibfield
  {journal} {\bibinfo  {journal} {Phys. Lett. B}\ }\textbf {\bibinfo {volume}
  {680}},\ \bibinfo {pages} {438} (\bibinfo {year} {2009})}\BibitemShut
  {NoStop}%
\bibitem [{\citenamefont {Sarriguren}(2012)}]{sarriguren2012JP}%
  \BibitemOpen
  \bibfield  {author} {\bibinfo {author} {\bibfnamefont {P.}~\bibnamefont
  {Sarriguren}},\ }\bibfield  {title} {\bibinfo {title} {Weak decay rates for
  waiting-point nuclei involved in the rp-process},\ }in\ \href@noop {} {\emph
  {\bibinfo {booktitle} {Journal of Physics: Conference Series}}},\ Vol.\
  \bibinfo {volume} {366}\ (\bibinfo {organization} {IOP Publishing},\ \bibinfo
  {year} {2012})\ p.\ \bibinfo {pages} {012039}\BibitemShut {NoStop}%
\bibitem [{\citenamefont {Nabi}(2012)}]{nabi2012AASS}%
  \BibitemOpen
  \bibfield  {author} {\bibinfo {author} {\bibfnamefont {J.-U.}\ \bibnamefont
  {Nabi}},\ }\bibfield  {title} {\bibinfo {title} {rp-process weak-interaction
  mediated rates of waiting-point nuclei},\ }\href@noop {} {\bibfield
  {journal} {\bibinfo  {journal} {Astrophysics and Space Science}\ }\textbf
  {\bibinfo {volume} {339}},\ \bibinfo {pages} {305} (\bibinfo {year}
  {2012})}\BibitemShut {NoStop}%
\bibitem [{\citenamefont {Nabi}\ and\ \citenamefont
  {B{\"o}y{\"u}kata}(2016)}]{nabi2016beta}%
  \BibitemOpen
  \bibfield  {author} {\bibinfo {author} {\bibfnamefont {J.-U.}\ \bibnamefont
  {Nabi}}\ and\ \bibinfo {author} {\bibfnamefont {M.}~\bibnamefont
  {B{\"o}y{\"u}kata}},\ }\bibfield  {title} {\bibinfo {title} {$\beta$-decay
  half-lives and nuclear structure of exotic proton-rich waiting point nuclei
  under rp-process conditions},\ }\href@noop {} {\bibfield  {journal} {\bibinfo
   {journal} {Nucl. Phys. A}\ }\textbf {\bibinfo {volume} {947}},\ \bibinfo
  {pages} {182} (\bibinfo {year} {2016})}\BibitemShut {NoStop}%
\bibitem [{\citenamefont {Nabi}\ and\ \citenamefont
  {B{\"o}y{\"u}kata}(2017)}]{Nabi2017AASS}%
  \BibitemOpen
  \bibfield  {author} {\bibinfo {author} {\bibfnamefont {J.-U.}\ \bibnamefont
  {Nabi}}\ and\ \bibinfo {author} {\bibfnamefont {M.}~\bibnamefont
  {B{\"o}y{\"u}kata}},\ }\bibfield  {title} {\bibinfo {title} {Nuclear
  structure and weak rates of heavy waiting point nuclei under rp-process
  conditions},\ }\href@noop {} {\bibfield  {journal} {\bibinfo  {journal}
  {Astrophysics and Space Science}\ }\textbf {\bibinfo {volume} {362}},\
  \bibinfo {pages} {9} (\bibinfo {year} {2017})}\BibitemShut {NoStop}%
\bibitem [{\citenamefont {Petrovici}\ \emph {et~al.}(2011)\citenamefont
  {Petrovici}, \citenamefont {Schmid},\ and\ \citenamefont
  {Faessler}}]{A.petrovici2011PPNP}%
  \BibitemOpen
  \bibfield  {author} {\bibinfo {author} {\bibfnamefont {A.}~\bibnamefont
  {Petrovici}}, \bibinfo {author} {\bibfnamefont {K.}~\bibnamefont {Schmid}},\
  and\ \bibinfo {author} {\bibfnamefont {A.}~\bibnamefont {Faessler}},\
  }\bibfield  {title} {\bibinfo {title} {Beyond mean field approach to the beta
  decay of medium mass nuclei relevant for nuclear astrophysics},\ }\href
  {https://doi.org/https://doi.org/10.1016/j.ppnp.2011.01.022} {\bibfield
  {journal} {\bibinfo  {journal} {Prog. Part. Nucl. Phys.}\ }\textbf {\bibinfo
  {volume} {66}},\ \bibinfo {pages} {287} (\bibinfo {year} {2011})},\ \bibinfo
  {note} {particle and Nuclear Astrophysics}\BibitemShut {NoStop}%
\bibitem [{\citenamefont {Petrovici}\ and\ \citenamefont
  {Andrei}(2015)}]{A.petrovici2015EPJA}%
  \BibitemOpen
  \bibfield  {author} {\bibinfo {author} {\bibfnamefont {A.}~\bibnamefont
  {Petrovici}}\ and\ \bibinfo {author} {\bibfnamefont {O.}~\bibnamefont
  {Andrei}},\ }\bibfield  {title} {\bibinfo {title} {Stellar weak interaction
  rates and shape coexistence for 68se and 72kr waiting points},\ }\href@noop
  {} {\bibfield  {journal} {\bibinfo  {journal} {Europ. Phys. J. A}\ }\textbf
  {\bibinfo {volume} {51}},\ \bibinfo {pages} {1} (\bibinfo {year}
  {2015})}\BibitemShut {NoStop}%
\bibitem [{\citenamefont {Petrovici}\ \emph {et~al.}(2019)\citenamefont
  {Petrovici}, \citenamefont {Mare}, \citenamefont {Andrei},\ and\
  \citenamefont {Meyer}}]{Petrovici_2019_PRC}%
  \BibitemOpen
  \bibfield  {author} {\bibinfo {author} {\bibfnamefont {A.}~\bibnamefont
  {Petrovici}}, \bibinfo {author} {\bibfnamefont {A.~S.}\ \bibnamefont {Mare}},
  \bibinfo {author} {\bibfnamefont {O.}~\bibnamefont {Andrei}},\ and\ \bibinfo
  {author} {\bibfnamefont {B.~S.}\ \bibnamefont {Meyer}},\ }\bibfield  {title}
  {\bibinfo {title} {Impact of $^{68}\mathrm{Se}$ and $^{72}\mathrm{Kr}$
  stellar weak interaction rates on $rp$-process nucleosynthesis and
  energetics},\ }\href {https://doi.org/10.1103/PhysRevC.100.015810} {\bibfield
   {journal} {\bibinfo  {journal} {Phys. Rev. C}\ }\textbf {\bibinfo {volume}
  {100}},\ \bibinfo {pages} {015810} (\bibinfo {year} {2019})}\BibitemShut
  {NoStop}%
\bibitem [{\citenamefont {Lau}(2018)}]{R.Lau2018NPA}%
  \BibitemOpen
  \bibfield  {author} {\bibinfo {author} {\bibfnamefont {R.}~\bibnamefont
  {Lau}},\ }\bibfield  {title} {\bibinfo {title} {Sensitivity tests on the
  rates of the excited states of positron decays during the rapid proton
  capture process of the one-zone x-ray burst model},\ }\href
  {https://doi.org/https://doi.org/10.1016/j.nuclphysa.2017.10.009} {\bibfield
  {journal} {\bibinfo  {journal} {Nucl. Phys. A}\ }\textbf {\bibinfo {volume}
  {970}},\ \bibinfo {pages} {1} (\bibinfo {year} {2018})}\BibitemShut {NoStop}%
\bibitem [{\citenamefont {Lau}(2020)}]{R.Lau2020MN}%
  \BibitemOpen
  \bibfield  {author} {\bibinfo {author} {\bibfnamefont {R.}~\bibnamefont
  {Lau}},\ }\bibfield  {title} {\bibinfo {title} {Sensitivity studies on the
  thermal beta+ decay and thermal neutrino decay rates in the one-zone x-ray
  burst model},\ }\href@noop {} {\bibfield  {journal} {\bibinfo  {journal}
  {Monthly Notices of the Royal Astronomical Society}\ }\textbf {\bibinfo
  {volume} {498}},\ \bibinfo {pages} {2697} (\bibinfo {year}
  {2020})}\BibitemShut {NoStop}%
\bibitem [{\citenamefont {Chen}\ and\ \citenamefont
  {Wang}(2024)}]{ZRChen_PLB2024}%
  \BibitemOpen
  \bibfield  {author} {\bibinfo {author} {\bibfnamefont {Z.-R.}\ \bibnamefont
  {Chen}}\ and\ \bibinfo {author} {\bibfnamefont {L.-J.}\ \bibnamefont
  {Wang}},\ }\bibfield  {title} {\bibinfo {title} {Stellar weak-interaction
  rates for $rp$-process waiting-point nuclei from projected shell model},\
  }\href {https://doi.org/https://doi.org/10.1016/j.physletb.2023.138338}
  {\bibfield  {journal} {\bibinfo  {journal} {Phys. Lett. B}\ }\textbf
  {\bibinfo {volume} {848}},\ \bibinfo {pages} {138338} (\bibinfo {year}
  {2024})}\BibitemShut {NoStop}%
\bibitem [{\citenamefont {Lv}\ \emph {et~al.}(2022)\citenamefont {Lv},
  \citenamefont {Sun}, \citenamefont {Fujita}, \citenamefont {Fujita},
  \citenamefont {Wang},\ and\ \citenamefont {Gao}}]{Lv_Cui_Juan_2022_PRC}%
  \BibitemOpen
  \bibfield  {author} {\bibinfo {author} {\bibfnamefont {C.-J.}\ \bibnamefont
  {Lv}}, \bibinfo {author} {\bibfnamefont {Y.}~\bibnamefont {Sun}}, \bibinfo
  {author} {\bibfnamefont {Y.}~\bibnamefont {Fujita}}, \bibinfo {author}
  {\bibfnamefont {H.}~\bibnamefont {Fujita}}, \bibinfo {author} {\bibfnamefont
  {L.-J.}\ \bibnamefont {Wang}},\ and\ \bibinfo {author} {\bibfnamefont
  {Z.-C.}\ \bibnamefont {Gao}},\ }\bibfield  {title} {\bibinfo {title} {Effect
  of nuclear deformation on the observation of a low-energy super-gamow-teller
  state},\ }\href {https://doi.org/10.1103/PhysRevC.105.054308} {\bibfield
  {journal} {\bibinfo  {journal} {Phys. Rev. C}\ }\textbf {\bibinfo {volume}
  {105}},\ \bibinfo {pages} {054308} (\bibinfo {year} {2022})}\BibitemShut
  {NoStop}%
\end{thebibliography}

%

\end{document}